# ICIPE-2024-0063
# A CONVOLUTIONAL MODEL FOR ESTIMATING THE JUNCTION TEMPERATURES OF SIC MOSFET TRANSISTORS

**Ali EL ARABI**
**Denis MAILLET**
**Nicolas BLET**
**Benjamin REMY**
Université de Lorraine, CNRS, LEMTA, Nancy F-54000, France
Ali.el-arabi@univ-lorraine.fr
Denis.maillet@univ-lorraine.fr
Nicolas.blet@univ-lorraine.fr
Benjamin.remy@univ-lorraine.fr

***Abstract.*** *The junction temperature is a very important parameter for monitoring power electronics converters based on MOSFET transistors. They offer the possibility of switching at relatively higher frequencies than other transistors like IGTBTs. However, the electrical parameters of MOSFETs are highly thermally dependent. The thermo-dependence of MOSFET electrical parameters is rarely taken into consideration when implementing control strategies, for many technological reasons, such as the difficulty of measuring the junction temperature. In practice, the junction temperature of transistors is inaccessible for direct measurement. The presence of a gel covering the chips, that provides electrical and thermal insulation, makes measurement by infrared thermography impossible. Furthermore, direct thermocouple measurement cannot be implemented due to the electromagnetic disturbances in the environment. Several researchers have attempted to correlate chip temperature with thermosensitive electrical parameters. In the present work, a thermal convolutive model has been developed to estimate the junction temperatures of two MOSFET transistors belonging to the same electronic circuit from external temperature measurements in two well-chosen locations (far away enough from the junction to avoid electromagnetic interference), using also the measured power dissipated on each chip. The thermal coupling between the two transistors has been considered in the form of mutual transmittances. The model was first calibrated using three-dimensional numerical simulations in COMSOL Multiphysics, followed by an experimental study. The results are very promising, illustrating the robustness of the convolutional model.*

## 1. INTRODUCTION

The energy transition and the decarbonization of energies require a better integration of renewable energies into the energy mix. But it is also a question of energy efficiency: doing the same thing with a lower energy consumption is mandatory. To achieve this, power electronic converters can reduce energy losses through optimal electrical control in order to maximize energy production (as an example MPPT: Maximum Power Point Tracker). In practice, the implementation of control strategies is based on several parameters such as voltage, current, switching frequency.... On the other hand, the transistors making up the electronic component carry relatively high currents (and therefore high electrical power) and switch at relatively high frequencies (about $100\ kHz$). This contributes to transistor overheating due to switching thermal losses and electrical conduction losses. This phenomenon is all the more important in silicon carbide-based MOSFET transistors, where both the electrical power passing through the component and its switching frequency are relatively high compared with other categories such as IGBTs and Si MOSFETs (Fig. 1).

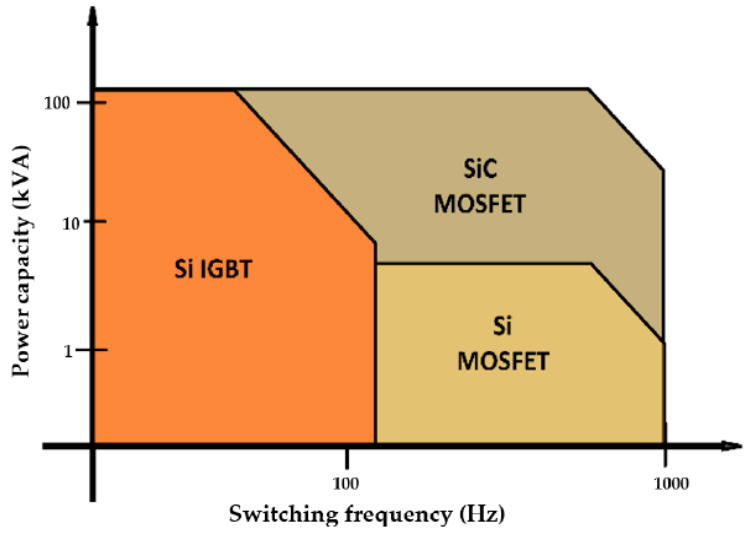

Figure 1. Operating power and frequency for MOSFETs and IGBTs (Isa et al. 2023)

MOSFETs offer a fairly high switching frequency and allow relatively high-power flow. However, their electrical parameters are highly thermosensitive (Dang et al.2016). This thermodependency makes temperature a key parameter to consider when implementing transistor control strategies. However, the temperatures to be monitored are those of the junctions, which are the highest. The problem is that these temperatures are not accessible for direct measurement because of several technological reasons. Temperature measurement by infrared thermography is not possible due to the presence of a gel covering the chips, that ensures both thermal insulation and dielectric rigidity of the medium in order to prevent the formation of electric arcs and thus the risk of a short circuit. This gel is opaque in the infrared range and cannot be removed in normal applications for security reasons, making its measurement by infrared thermography impossible. In addition, measurement by thermocouples or temperature-sensitive electrical sensors is not possible because they would be biased because of the harsh electromagnetic environment.

To overcome the problems mentioned above, we decided to set up a delocalized indirect temperature measurement system. This is a virtual junction temperature sensor enabling the junction temperature to be estimated from external temperature (and/or power) measurements. We have already presented in (El Arabi et al., 2023a and b) a semi-analytical thermal model for correlating the junction temperature with the temperature variations on the lower face and the thermal power dissipated on the top face (example of the pyramidal structure of such a MOSFET transistor in Fig. 2). The structure of the component is made of 5 parallelepiped blocks here, each block possibly having a multilayer structure. The associated "detailed model" takes the following form

$$\left[\widetilde{\widetilde{\theta}}_3^H \quad \widetilde{\widetilde{\theta}}_I \quad \widetilde{\widetilde{\theta}}_{II} \quad \widetilde{\widetilde{\theta}}_{III} \quad \widetilde{\widetilde{\theta}}_{IIII}\right]^T = H^{-1}\left(\widetilde{\widetilde{\theta}}_1^B - W\widetilde{\widetilde{\phi}}_3^H\right) \qquad (1)$$

where $\widetilde{\widetilde{\theta}}_3^H, \widetilde{\widetilde{\theta}}_1^B$ are the Fourier space harmonics of the time Laplace transform of temperature on the upper and lower face respectively, while $\widetilde{\widetilde{\phi}}_3^H$ are the corresponding harmonics of the Laplace transform of the heat flux on the upper face. $\widetilde{\widetilde{\theta}}_I$, $\widetilde{\widetilde{\theta}}_{II}$ and $\widetilde{\widetilde{\theta}}_{IIII}$ are the corresponding harmonics of Laplace temperature that are calculated over portions of horizontal intermediate levels that are not covered by another upper block. The only non-zero heat losses concern the lower horizontal face of the MOSFET, all the other free faces being adiabatic ones, except the top one where the heat dissipation on each chip is fully converted into corresponding heat fluxes.

The temperature variation at any point of the upper face is obtained by setting up a numerical inverse Laplace transform, then an analytical inverse Fourier Transform on $\widetilde{\widetilde{\theta}}_3^H$. The present semi analytical model has been verified by comparing its outputs to those of the commercial finite element code COMSOL Multiphysics (El Arabi et al., 2023b). It presupposes a perfect knowledge of the geometric parameters and thermophysical properties of the electronic components and heat exchange coefficients, as well as their variations with temperature. However, these data are not available for our modeling (manufacturing confidentialities or unknows). In addition, to simplify modelling, several assumptions were made, such as adiabaticity on lateral surfaces, and ignoring contact resistances and welds. This has limited the development of this model and led us to think of other models, such as the convolutional model presented in the present article.

The convolutional model, which is in fact a "reduced model" (Al Hadad et al, 2017), links the temperatures of the two junctions to the temperatures of the two temperature sensors (thermocouples) placed on the copper baseplate or anywhere inside the structure (Fig. 2). The development of this model involves three essential stages. The first is calibration, to identify the relationship between a junction temperature and a thermocouple temperature. Once these relationships (transmittances) have been identified, the same data set is used to attempt to estimate temperatures of junctions from temperatures of thermocouples - this is the verification stage. Once the first two steps have been successfully completed, the identified transmittances are used on a new data set to attempt to estimate the temperatures of junctions from the temperatures of thermocouples, as the validation stage.

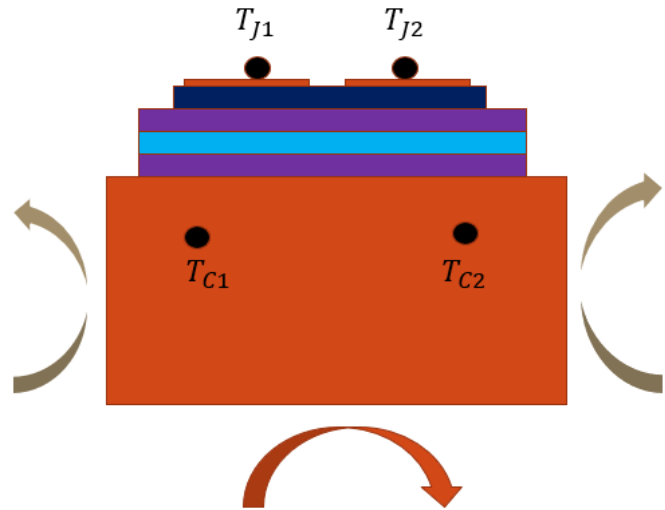

Figure 2. schematic diagram of the component used

To be able to conclude on the pertinence of this model reduction for later use in an inverse heat conduction problem, here the construction of a virtual junction temperature sensor, it must pass each of the three stages, otherwise it has to be changed (Beck et al., 1985).

In the following, we will present the convolutional model and the steps that lead to its development. We will then present the tests of pertinence of the convolutional model based on a three-dimensional numerical simulation in COMSOL Multiphysics that will be considered as the "detailed model". A noise study is also presented to conclude on the robustness of the convolutional model. Finally, the experimental setup and the validation of the convolutional model based on experimental results will be presented. It is also based on three steps (Identification, Verification and Validation) which will be carried out on real experimental data where the temperature of each junction is measured by infrared thermography and the temperatures of the based plate are measured by two thermocouples.

## 2. THE CONVOLUTIONAL MODEL

### 2.1 The convolutional model SISO (Single Input Single Output)

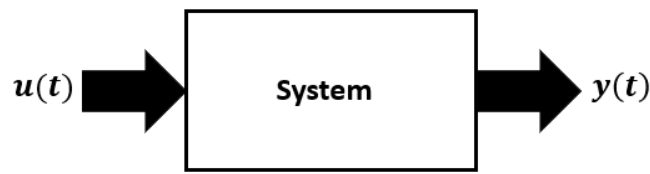

Figure 3. General scheme of a single input – single output dynamical system in heat transfer

For a system linking an input (source or pseudo-source: heat flux or temperature, $u(t)$ in Fig.3) to an output (consequence: heat flux or temperature at another point, $y(t)$ in the Fig.3), the causal relationship is defined by the convolution product (Al Hadad et al, 2019):

$$y(t) = (h * u)(t) \qquad (2)$$

Note that this reduced model, for a forced response to a transient excitation, is pertinent if the heat equation and its boundary conditions are linear, with time independent condition and with an separable excitation (writable as a product between its space support and its time intensity -Al Hadad et al., 2017).
The above equation could also be written in the integral form (Al Hadad et al., 2019):

$$y(t) = \int_0^{+\infty} u(t-\tau)h(\tau)\,d\tau \tag{3}$$

or in the discrete form:

$$y(t_i) = \Delta t \sum_{j=1}^{m} \tilde{u}_{i-j+1}\tilde{h}_j\,d\tau\ , \text{ for } i = 1,2\ldots\ldots n \tag{4a}$$

$$\text{where } \tilde{x}_j = \frac{1}{\Delta t}\int_{t_{j-1}}^{t_j} x(t)dt \approx \frac{1}{2}(x_j + x_{j-1}) \text{ for } x = h \text{ or } u \text{ where } x_j = x(t_j) \text{ with } t_j = j\Delta t \tag{4b}$$

Here $\tilde{x}$ is the average value of $x$ over an interval of length $\Delta t$. The above relationship could also be written in one of the two following matrix forms:

$$\mathbf{y} = \Delta t\ \mathbf{N}\ (\tilde{\mathbf{u}})\ \tilde{\mathbf{h}}\quad \text{or}\quad \mathbf{y} = \Delta t\ \mathbf{N}\ (\tilde{\mathbf{h}})\ \tilde{\mathbf{u}}\quad \text{with}\quad \tilde{\mathbf{x}} = \begin{bmatrix}\tilde{x}_1 & \tilde{x}_2 & \cdots & \tilde{x}_n\end{bmatrix}^T \text{ for } x = h \text{ or } u \tag{5a,5b}$$

Here $\mathbf{N}\ (.)$ is a function whose input is a column vector of size $n \times 1$ and output is a lower triangular Toeplitz matrix of size $n \times n$:

$$N(q) \equiv \begin{bmatrix} q_1 & 0 & 0 & 0 & 0 \\ q_2 & q_1 & 0 & 0 & 0 \\ q_3 & q_2 & q_1 & 0 & 0 \\ \vdots & \vdots & \vdots & \ddots & 0 \\ q_n & q_{n-1} & q_{n-2} & \cdots & q_1 \end{bmatrix} \text{ where } q = \begin{bmatrix} q_1 \\ q_2 \\ q_3 \\ \vdots \\ q_n \end{bmatrix} \tag{5c}$$

Equations (5a) to (5c) are used to estimate the transfer function $\tilde{h}$ averaged over intervals of length $\Delta t$ starting from the input and output values assumed to be known in the calibration step. The equation (5b) is used to estimate the output $y$ knowing the transfer function $h$ and input $u$. It can also be used to estimate the input from the output knowing the transfer function (source estimation problem).

It is important to note that equation (5a) gives access to the mean values of the transfer function and the input over each time step. The same model can be written with instantaneous values:

$$\mathbf{y} = M(\mathbf{u})\mathbf{h} \tag{6a}$$

$$\mathbf{y} = M(\mathbf{h})\mathbf{u} \tag{6b}$$

With:

$$M(\mathbf{x}) = \Delta t\ N^2(\mathbf{f})N(\mathbf{x}) \text{ and } \mathbf{f} = \frac{1}{2}[1\ 1\ 0\ \ldots\ \ldots\ 0]^T$$

The structure of model (6a) shows that the inversion of matrix $M\ (\mathbf{u})$ can give access to estimate the instantaneous values of the impulse response (vector $\mathbf{h}$) and that, once this value estimated, the inversion of matrix $M\ (\mathbf{h})$ in model (6b) can give access to the instantaneous input (vector $\mathbf{u}$). However, because both inversions would also require the inversion of matrix $N^2\ (\mathbf{f})$, which is very badly conditioned, instantaneous values can be difficult to estimated, even for noiseless input and output. Only time-averaged of the impulse response $\tilde{h}$ and of the input $\tilde{u}$ can be reached easily, i.e. by inversion of matrices of the form:

$$\tilde{M}(\mathbf{x}) = \Delta t N(\tilde{\mathbf{x}}) = \Delta t N(\mathbf{f})N(\mathbf{x}) = N(\mathbf{f}^+)M(\mathbf{x}) \text{ for } \mathbf{x} = \mathbf{h} \text{ or } \mathbf{u}$$

$$\text{where } N(\mathbf{f}^+) = (N(\mathbf{f}))^{-1}\ \text{ with }\ \mathbf{f}^+ = 2[1\ -1\ 1\ -1\ \cdots\ (-1)^{n-1}]^T \tag{6c}$$

However, in this work we have tried to find local values of transmittances and temperatures have been estimated using (6a & 6b).

## 2.2 The convolutional model MIMO (Multiple Inputs Multiple Outputs)

Our case study is a MIMO configuration with multiple inputs and multiple outputs. In the present work, a system with two inputs and two outputs will be presented (Fig.2). The input can be a real source of heat like the thermal power dissipated in each chip ($P_1$ for the first transistor and $P_2$ for the second transistor), or a pseudo source such as another temperature like the junction temperature ($T_{J1}$ for the first transistor and $T_{J2}$ for the second transistor) which is itself a consequence of the thermal power dissipated in the chips. The output is the temperature of a sensor placed further away from the junction ($T_{C1}$ for the sensor below the first transistor and $T_{C2}$ for the sensor below the second transistor).

This section is purely illustraCive of the convolutional model with two inputs and two outputs. To do this, and to keep the general aspect of the subject, we will note $y_1$ and $y_2$ the two outputs, and $u_1$ $et$ $u_2$ the two inputs. The relationship between outputs and inputs is given by Duhamel's theorem and the principle of superposition:

$$y_1(t) = (h_{11} * u_1)(t) + (h_{12} * u_2)(t) \tag{7a}$$

$$y_2(t) = (h_{21} * u_1)(t) + (h_{22} * u_2)(t) \tag{7b}$$

The previous equations could also be written in the integral form:

$$y_1(t) = \int_0^{+\infty} u_1(t-\tau)h_{11}(\tau)\,d\tau + \int_0^{+\infty} u_2(t-\tau)h_{12}(\tau)\,d\tau \tag{8a}$$

$$y_2(t) = \int_0^{+\infty} u_1(t-\tau)h_{21}(\tau)\,d\tau + \int_0^{+\infty} u_2(t-\tau)h_{22}(\tau)\,d\tau \tag{8b}$$

The previous relationships could also be written under matrix form, adopted to estimate the transfer functions from inputs and outputs instantaneous values:

$$\boldsymbol{y_1} = M(\boldsymbol{u_1})\boldsymbol{h_{11}} + M(\boldsymbol{u_2})\boldsymbol{h_{12}} \tag{9a}$$

$$\boldsymbol{y_2} = M(\boldsymbol{u_1})\boldsymbol{h_{21}} + M(\boldsymbol{u_2})\boldsymbol{h_{22}} \tag{9b}$$

These relationships could be written differently to suit the estimation of sources from outputs and transfer functions:

$$\boldsymbol{y_1} = M(\boldsymbol{h_{11}})\boldsymbol{u_1} + M(\boldsymbol{h_{12}})\boldsymbol{u_2} \tag{10a}$$

$$\boldsymbol{y_2} = M(\boldsymbol{y_{21}})\boldsymbol{u_1} + M(\boldsymbol{h_{22}})\boldsymbol{u_2} \tag{10b}$$

The previous equations could be written in the following matrix form:

$$\begin{bmatrix} \boldsymbol{y_1} \\ \boldsymbol{y_2} \end{bmatrix} = P(\boldsymbol{h_1}, \boldsymbol{h_2}) \begin{bmatrix} \boldsymbol{u_1} \\ \boldsymbol{u_2} \end{bmatrix} \tag{11}$$

With :

$$P(\boldsymbol{x}, \boldsymbol{y}) = \begin{bmatrix} M(\boldsymbol{x_1}) & M(\boldsymbol{x_2}) \\ M(\boldsymbol{y_1}) & M(\boldsymbol{y_2}) \end{bmatrix}; \boldsymbol{x} = [\boldsymbol{x_1}, \boldsymbol{x_2}] \; and \; \boldsymbol{y} = [\boldsymbol{y_1}, \boldsymbol{y_2}]$$

To estimate the four impulse responses (transfer functions in the Laplace domaine), two sets of data ( which can come from two numerical simulations or from a real experimental system) are required. Each set of data correspond to a single active input $\boldsymbol{u}$, while the second one is kept at a zero value. Impulse responses can be estimated from the following systems:

$$\begin{bmatrix} \boldsymbol{y_1^{Data\,set\,1}} & \boldsymbol{y_1^{Data\,set\,2}} \end{bmatrix}^T = P(\boldsymbol{u^{Data\,set\,1}}, \boldsymbol{u^{Data\,set\,2}})H_1 \tag{12a}$$

$$\begin{bmatrix} \boldsymbol{y_2^{Data\,set\,1}} & \boldsymbol{y_2^{Data\,set\,2}} \end{bmatrix}^T = P(\boldsymbol{u^{Data\,set\,1}}, \boldsymbol{u^{Data\,set\,2}})H_2 \tag{12b}$$

with $\boldsymbol{u^{Data\,set}} = [\boldsymbol{u_1^{Data\,set}}, \boldsymbol{u_2^{Data\,set}}]$; $H_1 = [\boldsymbol{h_{11}}\;\boldsymbol{h_{12}}]^T$ ; $H_2 = [\boldsymbol{h_{21}}\;\boldsymbol{h_{22}}]^T$

Note here that the vectors $\boldsymbol{h_{11}}, \boldsymbol{h_{22}}$ represent the inherent impulse responses relating each output *i* to the input *i* (closest to each other). The $\boldsymbol{h_{12}}, \boldsymbol{h_{21}}$ functions represent coupling transfer functions, i.e. the effect of source 1 on output 2 for $\boldsymbol{h_{21}}$ and vice versa for $\boldsymbol{h_{12}}$.

## 3. ESTIMATION OF THE JUNCTION TEMPERATURES OF TWO TRANSISTORS USING THE CONVOLUTIONAL MODEL
### 3.1 Case study presentation

The studied component is shown in Fig. 2 in section 1. This device contains two silicon carbide-based MOSFET transistors. The electronic configuration of the system (the two transistors mounted on the same electric arm) does not allow both transistors to operate at the same time, but they can operate in phase opposition. The aim of the present work is to estimate the junction temperatures ($T_{J1}$ and $T_{J2}$) from the temperature values of the two sensors ($T_{C1}$ and $T_{C2}$). Another objective is to be able to estimate junction temperatures from the value of the thermal power dissipated on each transistor chip. For the numerical simulation, the convection exchange coefficient on the lower face was taken equal to $2500\ W/(m^2.K)$. The value for the side walls was taken equal to $100\ W/(m^2.K)$. The power dissipated in each component in the calibration step was taken equal to $100\ W/m^2$. To better track the rapid temperature changes at the junctions, a time step of $10\ ms$ was used for the time discretization with a simulation time of 15 seconds.

Table 1. Geometric dimensions and thermophysical property values

|  | Dimensions ($l \times w \times h$) | $\rho(kg/m^3)$ | $c_p(J/(kg.K))$ | $\lambda(W/(m.K))$ |
|---|---|---|---|---|
| copper heat sink | $36.0 \times 35.0 \times 15.0\ mm$ | 8900 | 390 | 400 |
| electronic board | $36.0 \times 35.0 \times 1.0\ mm$ | 6400 | 530 | 240 |
| transistor (Silicon carbide) | $1.25 \times 1.25 \times 0.10\ mm$ | 3000 | 650 | 130 |

We will also test the effect of the transmittance estimation method on the final result in terms of junction temperatures. In fact, the relationship between junction temperature and sensor temperatures can be identified directly. In this case, junction temperatures are considered as inputs (pseudo-source) and sensor temperatures as outputs. Transmittances linking junction temperatures to sensor temperatures can be identified directly using equations (12a & 12b).

### 3.2 Estimating junction temperatures from thermocouple temperatures

The relationship between the temperatures elevation ($\theta(t) = T(t) - T(0)$) of the two sensors and the two junction temperatures is given by:

$$\theta_{C1}(t) = (w_{11} * \theta_{J1})(t) + (w_{12} * \theta_{J2})(t) \tag{13a}$$

$$\theta_{C2}(t) = (w_{21} * \theta_{J1})(t) + (w_{22} * \theta_{J2})(t) \tag{13b}$$

then under matrix form:

$$\boldsymbol{\theta_c} = P(\boldsymbol{W_1}, \boldsymbol{W_2})\boldsymbol{\theta_J} \tag{13c}$$

with $\boldsymbol{\theta_c} = [\theta_{c1}\ \theta_{c2}]^T$ ; $\boldsymbol{\theta_J} = [\theta_{J1}\ \theta_{J2}]^T$ ; $\boldsymbol{W_1} = [w_{11}\ w_{12}]^T$ ; $\boldsymbol{W_2} = [w_{21}\ w_{22}]^T$
and $w_{ij}$ are the transmittances linking the sensors to the junctions.

The above relationship allows to estimate junction temperatures from sensor temperatures by inversion of the transmittance matrix ($P(\boldsymbol{W_1}, \boldsymbol{W_2})$). The four transmittances can be identified using the two systems:

$$[\theta_{c1}^{Data\ set\ 1}\ \theta_{c1}^{Data\ set\ 2}]^T = P(\theta_J^{Data\ set\ 1}, \theta_J^{Data\ set\ 2})\boldsymbol{W_1} \tag{14a}$$

$$[\theta_{c2}^{Data\ set\ 1}\ \theta_{c2}^{Data\ set\ 2}]^T = P(\theta_J^{Data\ set\ 1}, \theta_J^{Data\ set\ 2})\boldsymbol{W_2} \tag{14b}$$

The determinant of the matrices is equal to zero and the condition number is very large. This is why the singular value decomposition method with truncation (TSVD) will be used to invert these matrices (Al Hadad et al. 2013). Note that two simulations are necessary under COMSOL Multiphysics to estimate transmittances, where only one chip is activated in each simulation to investigate coupling effects.

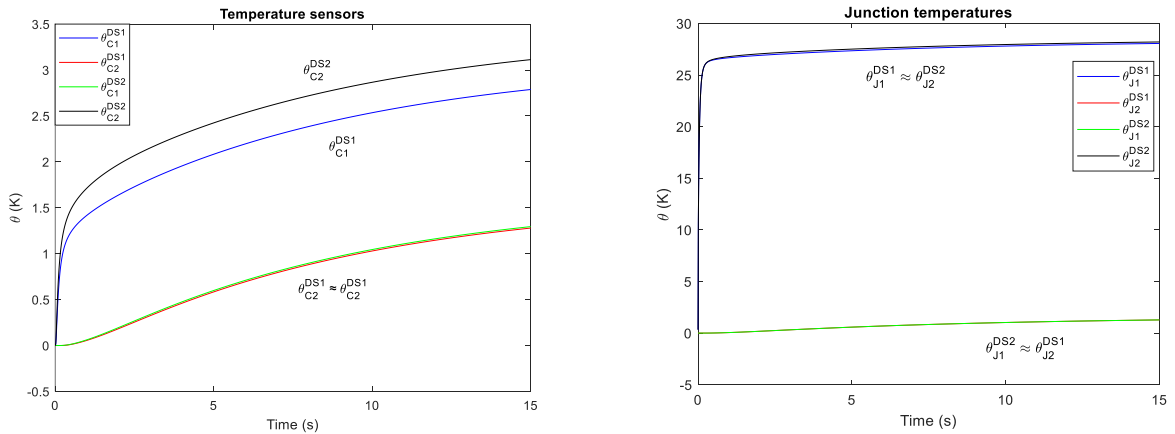

Figure 4. Temperature rises at junctions and sensors for both data sets (DS1: $P_1 = 100W/cm^2, P_2 = 0$; DS2: $P_1 = 0\ W/cm^2, P_2 = 100\ W/cm^2$) in the direct problem configuration

In the first simulation, a power step is imposed on the first chip and the second chip is switched off. In the second simulation, a power step is applied to the second chip and the first is switched off. The temperatures obtained in the two simulations are shown in Fig.4. The four transmittances identified are illustrated in Fig. 5.

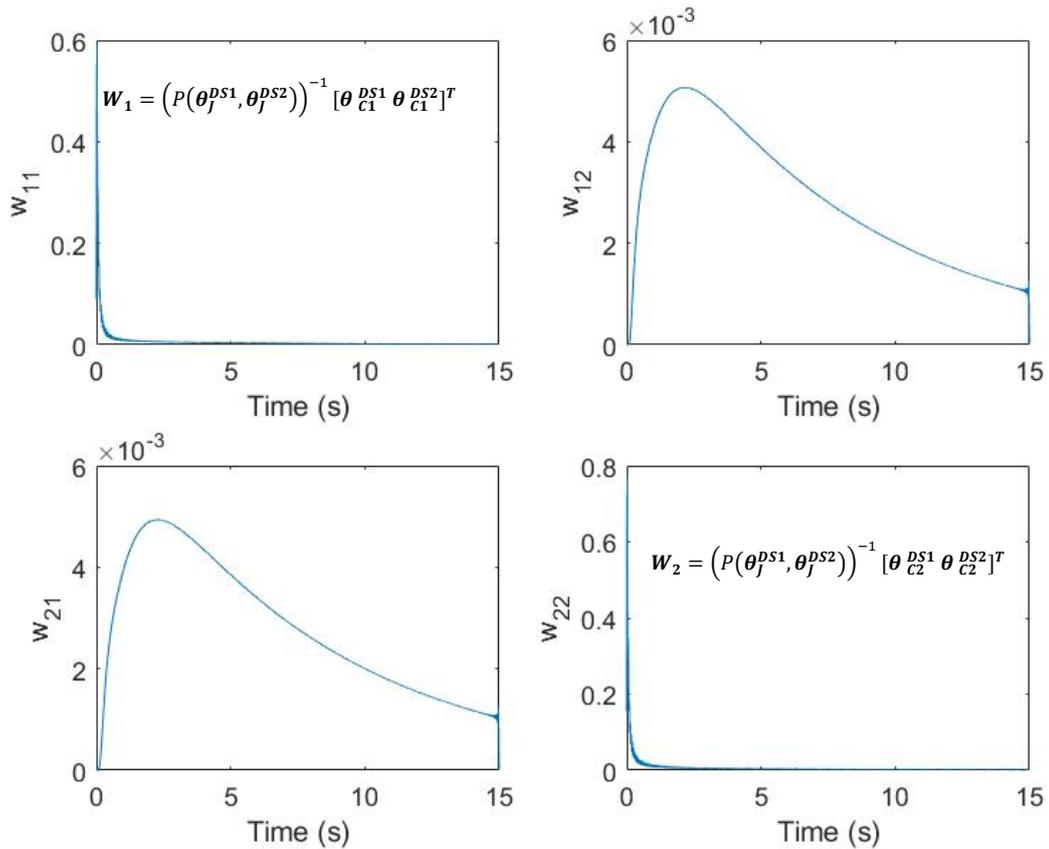

*Figure 5. transmittances identified from 3D numerical simulation*

The geometry appears to be symmetrical through the median plane. However, a difference between the maximum value of $w_{11}$ and that of $w_{22}$ can be observed. This is justified by the fact that the distance between sensor 2 and junction 2 is smaller than that between sensor 1 and the junction 1 (representative on the experimental bench). This comment can also be supported by the fact that the equilibrium temperature at junction 2 for the second data set is higher than that at junction 1 for the first data set. The transmittances are now used to estimate the junction temperatures from the temperatures of the two thermocouples, using Eq. (14). Figure 6 illustrates that the convolutional model is able to recover the detailed simulation results with a very good fit, for both junctions and sensors.

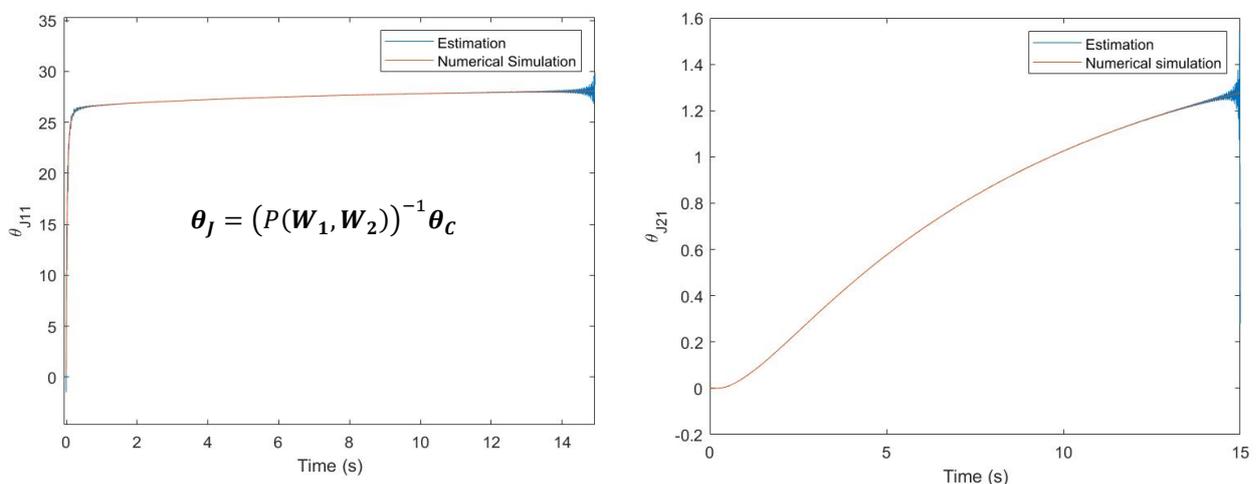

*Figure 6. Temperature variations of the two junctions for the calibration data set one*

For the "validation" dataset, a square-wave power signal was imposed in each chip with a period of five seconds and a duty cycle of 50%. The results of the numerical simulation and the temperature rise of the two junctions estimated from the temperature rises in the two thermocouples are shown in Fig. 7. Good agreement between numerical simulation and convolutional model results was observed. This means that the validation stage was successful and that the model can be used for other cases. The fluctuations observed in the estimated temperature rises are due to the use of the TSVD method to invert the matrix $P(W_1, W_2)$ : 1500 points in time are simulated; the singular matrix is of size $1200 \times 1200$, which represents a truncation of 20%.

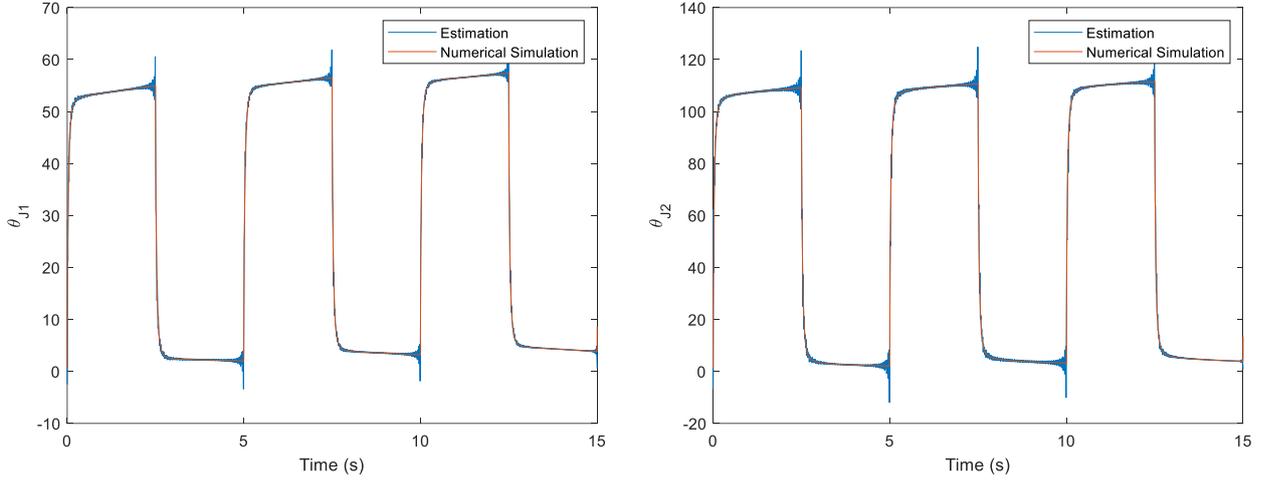

Figure 7. Temperature variations of the two junctions for the validation data set ($P_1 \neq 0$ and $P_2 \neq 0$ ; $\max(P_2) = 2 \times \max(P_1)$ )

## 4. EXPERIMENTAL STUDY

### 4.1 Presentation of the experimental bench

The experimental bench (Fig. 8) is based on a component containing two SiC MOSFET transistors. The component is placed on a copper block, where two thermocouples have been inserted in it. The assembly is placed on a cooling unit composed of a finned diffuser and a fan. Junction temperatures are measured by infrared thermography. To enable this, the gel covering the chips was chemically removed using acetone. The chips were painted with aerosols in black to overcome emissivity problems and improve the power of the thermal signal received by the camera.

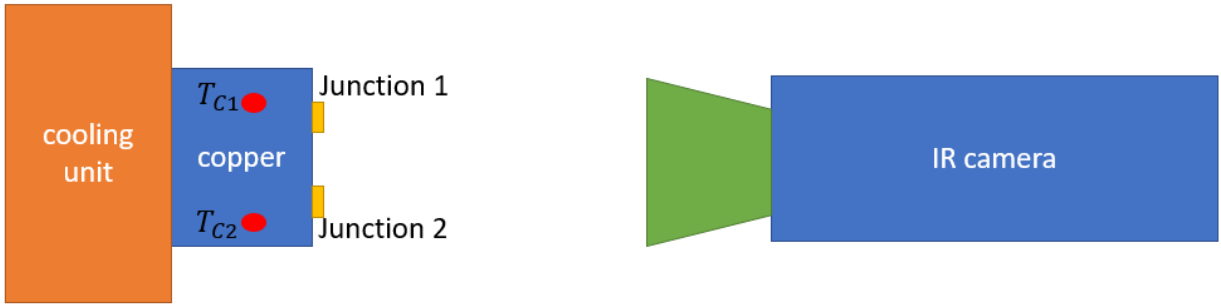

Figure 8. Diagram of the experimental bench

It should be noted that the simulations were carried out for a 15 s duration whereas experimental durations were much larger (200 s). This is due to technological limitations (maximum temperature supported by the component, etc.) and also to avoid the problem of measurement noise, which is very significant. Thus, the time step used in numerical simulation ($\Delta t = 0.01\ s$) is ten times smaller than that used experimentally ($\Delta t = 0.1\ s$). This choice was made to avoid the noise problem.

### 4.2 Experimental identification of transmittances

Calculating the temperature rise experimentally is not easy. So, in the present work the acquisition has been started five seconds before starting the power step. The mean value of the signal acquired before the start of the power step is used as the reference temperature. Using Eq. (15) with the TSVD method, the four transmittances can be determined (Fig. 9). We observe the same shape as in the numerical simulation case, with great similarity between $w_{12}$ and $w_{21}$. The amplitude of $w_{11}$ is greater than $w_{22}$, which can be explained by the proximity of thermocouple number 1 to junction 1. The fluctuations observed are the combined effect of measurement noise and the use of the TSDV method: 2000 points in time are used and the singular matrix is of size $1200 \times 1200$, which represents a truncation of 40%. Here, the identified transmittances seem less precise from those obtained in the numerical simulation. This can be explained by several factors, such as the difference between the numerical model and the real experiment. But one of the main reasons is related to measurement noise at the sensor and junction temperatures, and the use of TSVD with a fairly large truncation. However, despite the difficulty of estimating transmittances, we were able to find temperatures on both components.

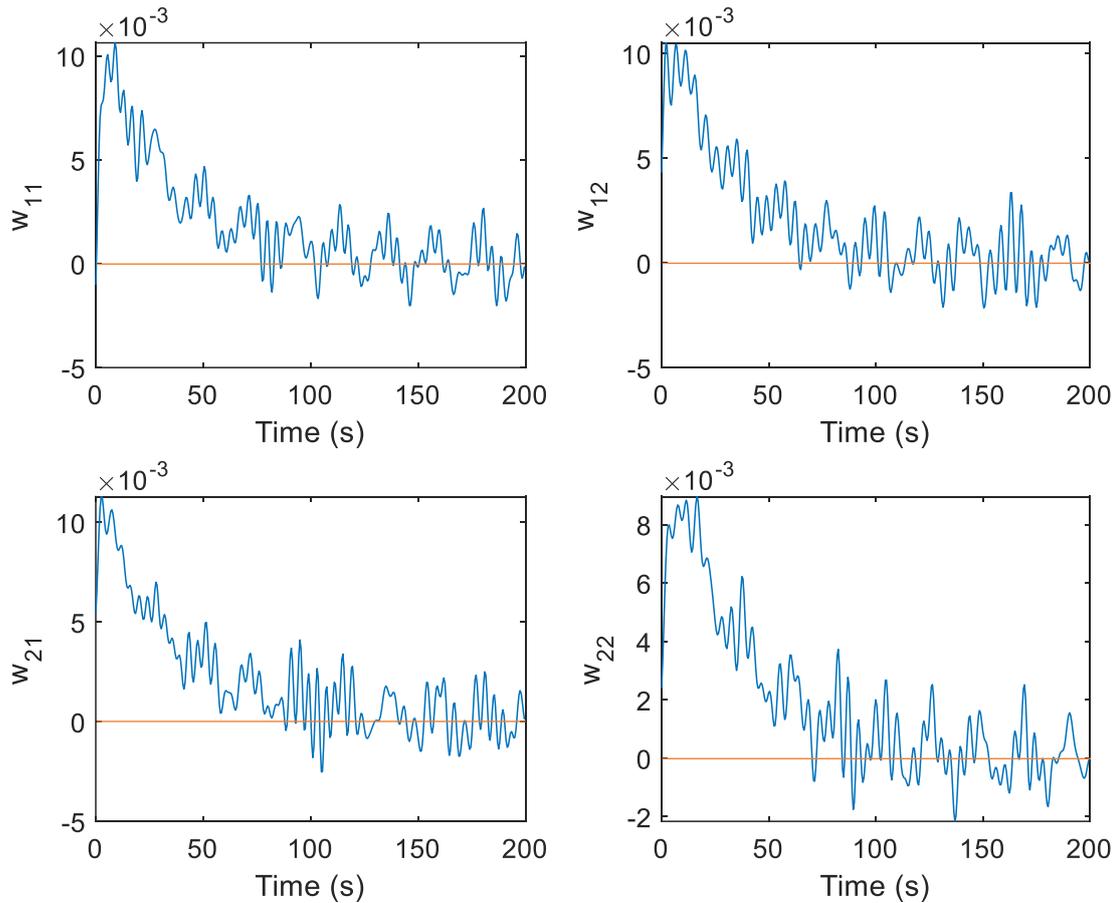

Figure 9. The transmittances identified experimentally

**4.3 The verification step**

The transmittances, identified and presented in the previous section, are now used to attempt to reproduce the temperature rises of the two junctions from the temperature rises in the two sensors. Figure 10 shows that the convolutional model is able to recover quite well the temperature rise profile of the two junctions, even if the estimation deviates slightly more from the measurements when steady state is reached, because of the higher relative uncertainty in transmittance values. Nevertheless, it can be noted that the calculation of the area under the curve is almost the same between the experimental and the convolutional cases, showing that the estimation is unbiased.

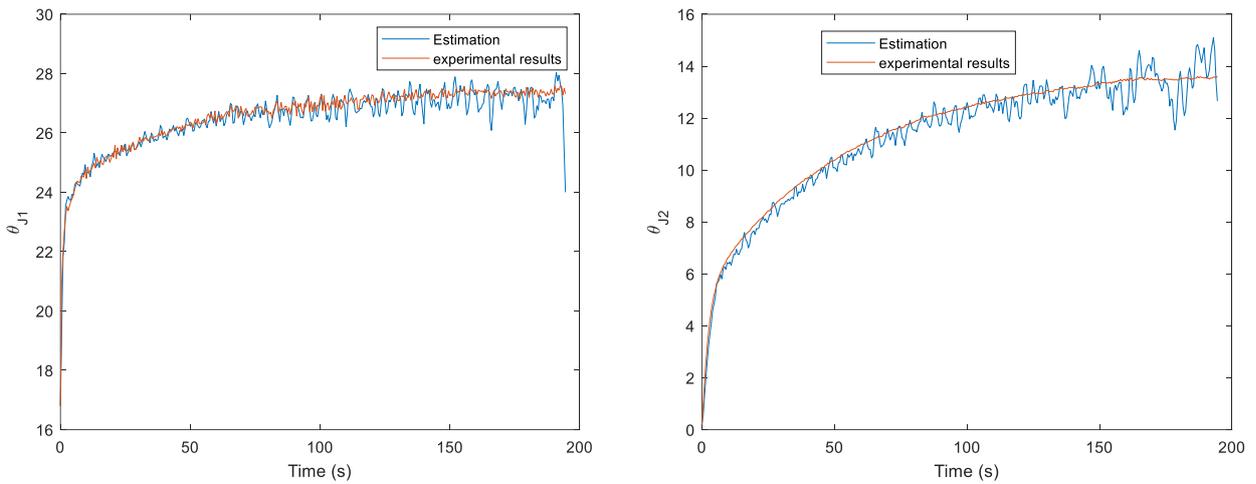

Figure 10. Temperature variations of the two junctions for the calibration data set

**4.4 The validation step**

For a new data set, the convolutional model is able to find the temperature rise (Fig. 9) at junction 2 from the temperature rise measured by the two thermocouples, which allows us to perform an early validation of the convolutional model.

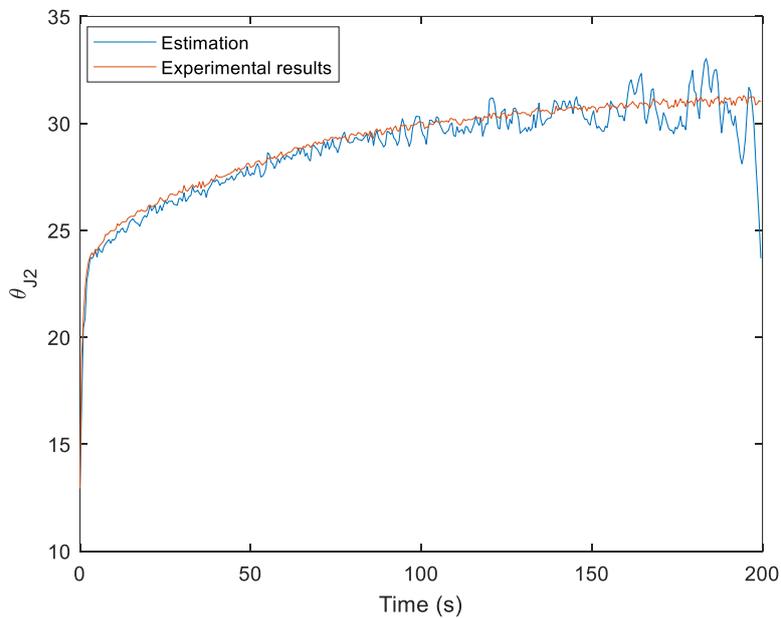

Figure 11. temperature difference at the second junction in the validation case

**5. CONCLUSIONS**

The results obtained by the convolutional model presented here are promising, since the estimation of two junction temperatures seems to be unbiased, even if signal noise problems are significant on the values of identified transmittances. Furthermore the convolutional model is costly in terms of computational resources (computation time of the order of one minute), but it is far better than a full numerical model (computation time of a few hours). An autoregressive model with an exogenous variable (ARX) may be a less costly solution, as fewer parameters need to be identified. The present study also enabled us to validate the approach experimentally.

# 7. RESPONSIBILITY NOTICE

The authors are the only responsible for the material included in this paper.